 \documentclass[12pt]{ronbun}
\usepackage{amsmath,amssymb}
\usepackage{graphicx,color}
\usepackage{cite}
\usepackage{bm}
\usepackage{dcolumn}
\newcommand{\bea}{\begin{eqnarray}}
\newcommand{\eea}{\end{eqnarray}}
\newcommand{\nnb}{\nonumber}
\newcommand{\bref}[1]{(\ref{#1})}
\newcommand{\nn}{\nonumber}

\newcommand{\del}{\delta}
\newcommand{\ra}{\rangle}
\newcommand{\la}{\langle}

\renewcommand{\th}{\theta}%
\numberwithin{equation}{section}

\begin{document}
\begin{flushright}
\parbox{4.2cm}
{KEK-TH-970 \hfill \\
{\tt hep-th/0407044}
 }
\end{flushright}

\vspace*{1.1cm}

\begin{center}
 \Large\bf Classical Integrability and Super Yangian  \\
of Superstring on AdS$_5$ $\times$ S$^5$ 
\end{center}
\vspace*{1.5cm}
\centerline{\large Machiko Hatsuda$^{\ast\dagger a}$ and Kentaroh
Yoshida$^{\ast b}$}
\begin{center}
$^{\ast}$\emph{Theory Division, High Energy Accelerator Research 
Organization (KEK),\\
Tsukuba, Ibaraki 305-0801, Japan.} 
\vspace*{0.5cm}
\\
$^{\dagger}$\emph{Urawa University, Saitama 336-0974, Japan}

\vspace*{1cm}
$^{a}$mhatsuda@post.kek.jp
~~~~
 $^{b}$kyoshida@post.kek.jp
\end{center}

\vspace*{1cm}

\centerline{\bf Abstract}
  
\vspace*{0.5cm}

We discuss a classical integrability in the type IIB string theory on
the AdS$_5\times$ S$^5$ background.  By using the Roiban-Siegel
formulation of the superstring on the AdS$_5\times $S$^5$\,, we
carefully treat the Wess-Zumino term and the constraint conditions
intrinsic to the supersymmetric case, and construct explicitly non-local
charges for a hidden infinite-dimensional symmetry. The existence of the
symmetry is shown by Bena-Polchinski-Roiban. Then the super Yangian
algebra is calculated.  We also show the Serre relation ensuring the
structure of the Hopf algebra. In addition, the classical integrability is
discussed by constructing the Lax pair and the transfer matrix.

\vfill 
\noindent {\bf Keywords:}~~{\footnotesize Integrability, Yangian
symmetry, spin chain, AdS/CFT correspondence} 

\thispagestyle{empty}
\setcounter{page}{0}

\newpage 

\section{Introduction}

One of the most important subjects in string theory is to test the
AdS/CFT correspondence \cite{M,GKP,W} beyond supergravity (i.e.\ BPS
sectors).  In the analysis at almost BPS region, a great development was
made by Berenstein-Maldacena-Nastase (BMN) \cite{BMN}. They presented
the AdS/CFT correspondence at string theoretical level by using the
Penrose limit and the exact solvability of the pp-wave string
theory. The pp-wave string states and their energies correspond to the
BMN operators and the full conformal dimensions (including the anomalous
dimensions) in the $\mathcal{N}=4$ super Yang-Mills (SYM) side,
respectively. After the BMN's work, an approach beyond the BMN limit was
proposed via the semiclassical quantization of string theory around a
classical solution (sigma model approach) \cite{GKP2, FT} instead of the
BPS null geodesics used in \cite{BMN}. The correspondence of
semiclassical string states and composite operators in $\mathcal{N}=4$
SYM is precisely formulated in \cite{BFST}, and it implies that the
AdS/CFT correspondence at non-BPS region.

In the SYM theory side  Minahan and Zarembo found the integrable
structure by studying the composite operators \cite{MZ}. 
They discussed the $SO(6)$ sector in the $\mathcal{N}=4$ SYM and studied
the operator mixing of local composite operators at one-loop planer
level by using the standard perturbation theory with respect to the 't
Hooft coupling $\lambda$ in the large $N$ limit. Then they found that
one-loop planer anomalous dimension matrix can be expressed as a
Hamiltonian of the $SO(6)$ quantum Heisenberg spin chain. This spin
chain can be solved by using the algebraic Bethe ansatz.  The
eigen-values and eigen-states give the anomalous dimensions and the
corresponding operators.  Further developments have been done, for
example, these operators include chiral operators, Konishi operators,
and the complete one-loop dilatation operators found by Beisert
\cite{B}. 

On the other hand, the integrable structure in the string theory side,
which is related to non-local charges in two-dimensional sigma models,
was discussed \cite{MSW}.  In 1978, L$\ddot{\rm u}$scher and Pohlmeyer
\cite{LP} found an infinite set of conserved non-local charges in
two-dimensional classical non-linear sigma models (NLSM$_2$).  The
existence of classical non-local charges leads to quantum non-local
charges which give constraints on the S-matrix and factorize it. Hence,
non-local charges are intimately connected to the classical and quantum
integrabilities of the theories.  These charges are related to the
Kac-Moody algebra \cite{Dolan} and generate the Drinfel'd Yangian
algebra \cite{Yangian} in the work \cite{Bernard} (For review of Yangian
symmetry, see \cite{Yangian:review}).  That is, the non-local charges
give a quantum field realization of the Yangian.

The existence of classical integrability and non-local charges in the
type IIB string theory on the AdS$_5\times $S$^5$ was suggested in the
early work \cite{MSW} by regarding the $S^5$ part as a NLSM. The bosonic
coset, $ {SO(2,4)}/{SO(4,1)} \times {SO(6)}/{SO(5)} = {\rm AdS}_5 \times
{\rm S}^5\,,$ is a symmetric space and so the AdS$_5\times$ S$^5$ valued
NLSM has non-local charges.  Contrasting to that for a symmetric coset
space the flat current and the non-local charges are studied well for
example in \cite{AAR}, for a non-symmetric coset space general arguments
have not been established yet so far.  A coset space for a superstring
system is not a symmetric space which reflects the superalgebra,
$\{Q,Q\}\sim P$.  For such a non-symmetric coset space $
{PSU(2,2|4)}/({SO(1,4)\times SO(5)})$\,, Bena-Polchinski-Roiban
\cite{BPR} showed the existence of non-local charges by constructing the
flat current based on the Maurer-Cartan equation in the simpler notation
of \cite{RS} and the field equations obtained in \cite{MT}.  The
Green-Schwarz (GS) superstring on the AdS$_5\times$ S$^5$ contains the
Wess-Zumino (WZ) term and the $\kappa$-symmetry.  Unlike the sigma models
with the supergroup target spaces \cite{BZV} the GS action does not
contain fermionic currents in the kinetic term but only in the WZ term,
so the superstring system has the fermionic constraints which must be
taken into account.

Dolan-Nappi-Witten also explained the appearance of spin chain
Hamiltonian in the SYM side from the viewpoint of the Yangian symmetry
\cite{DNW} which is assumed to be related to the non-local charges
found by Bena-Polchinski-Roiban \cite{BPR}. The classical/quantum
integrabilities in string/gauge theories are expected to play an
important role in a study of the AdS/CFT correspondence at non-BPS
region.

The next step in a study of the classical integrability of
AdS-superstring is to construct explicitly non-local charges by starting
from a definite action of AdS-superstring.  
In this paper we discuss a
classical integrability in type IIB string theory on the AdS$_5\times$
S$^5$ background. 
Firstly, we construct explicitly non-local charges, which are $\kappa$-invariant, 
by
using the Roiban-Siegel formulation \cite {RS} of superstring on the
AdS$_5 \times $S$^5$\,.  This formulation is more suitable to discuss
non-local charges rather than the Metsaev-Tseytlin formulation
\cite{MT}, since (1) the WZ term is bilinear of the left-invariant (LI)
currents, (2) the global symmetry is realized linearly and (3) the
matrix valued coordinates give simpler expression of the LI currents
rather than using an exponential parameterization.  Next, the super
Yangian algebra is presented by calculating the Poisson bracket of the
non-local charges.  The resulting algebra represents a $GL(4|4)$ super
Yangian algebra.  In that time the Serre relation is also shown. In
addition, we discuss the classical integrability of the AdS-superstring
theory in the Roiban-Siegel formulation by considering the Lax pair and
transfer matrix. Following the standard discussion, the Poisson bracket
of the transfer matrices gives a classical $r$-matrix, which satisfies the
classical Yang-Baxter equation.

This paper is organized as follows: In section 2 we briefly review
the Roiban-Siegel formulation of superstring on the 
AdS$_5\times$ S$^5$\,, 
and we obtain the conserved flat currents which are the right-invariant currents. 
In section 3 we explicitly construct an infinite set of
non-local charges.  Then we discuss the Yangian algebra generated by the
first two of them. The Serre relation is also discussed.  Moreover, we
show the Lax pair and transfer matrix of AdS-superstring, and then 
the classical integrability is discussed. 
Section 4 is devoted to a conclusion and discussions.

\section{Roiban-Siegel Formulation of Superstring on 
the AdS$_5$ $\times$ S$^5$} 

In this section we will briefly review the Roiban-Siegel formulation
\cite{RS} of the type IIB superstring on the AdS$_5\times$ S$^5$ background.

In the well-known approach of Metsaev-Tseytlin to construct the action
of the type IIB string on the AdS$_5\times$ S$^5$\,, a coset superspace
construction is used.  In this framework, strings propagate on the
superspace:
\begin{eqnarray}
\frac{PSU(2,2|4)}{SO(1,4)\times SO(5)}\,,
\end{eqnarray}
which has as even part the AdS$_5\times$ S$^5$ geometry.  The exponential
parameterization of the coset representation, however, is very
complicated and seems to be unsuitable to study non-local charges or
classical integrability in which we are interested in this paper.

On the other hand, the Roiban-Siegel formulation is appropriate to our
consideration. The starting point of this formulation is to simplify the
coset superspace with Wick rotations\footnote{In the following
discussion, the group theoretical features are only concerned, and hence
the signature is not so important.  In any case, we can Wick rotate
back.} as follows:
\begin{eqnarray}
\frac{PSU(2,2|4)}{SO(1,4)\times SO(5)} \quad \rightarrow \quad  
\frac{PSL(4|4)}{(Sp(4))^2} \quad \rightarrow  \quad 
\frac{GL(4|4)}{(Sp(4)\times GL(1))^2}\,. 
\end{eqnarray}
The Lie-algebra identifications: $SU(4) \cong SL(4)$ and $SO(5) \cong
Sp(4)$ lead to the first rewriting and the relaxation of both the $P$
and the $S$ constraints means the second one.  The two $GL(1)$'s can be
chosen to act separately on the upper-left and lower-right blocks. It
may be possible to use the $PSL(4|4)$ form rather than $GL(4|4)$\,. The
$PSL(4|4)$\,, however, does not have the representation in Mat$(4|4)$\,,
and so it is still inconvenient. 

The coset elements $Z_M^{~A}$\,, 
\[
 {\rm i.e.,} \qquad Z_M{}^{A} = 
\begin{pmatrix}
Z_m{}^{a} & Z_m{}^{\bar{a}} \\
Z_{\bar{m}}{}^{a} & Z_{\bar{m}}{}^{\bar{a}}
\end{pmatrix}
\quad 
 \in \frac{GL(4|4)}{(Sp(4)\times GL(1))^2}\,,
 \]
transform in the defining representation of the superconformal group.
The index $M$ is acted upon by the global coordinate transformation
$GL(4|4)$ while the local $(Sp(4)\times GL(1))^2$ (i.e., local Lorentz
($Sp(4)$'s) and dilatation ($GL(1)$'s) acts on the $A$ indices. 
The indices $m$ $(a)$ and $\bar{m}$ $(\bar{a})$ 
are $GL(4)$'s ($Sp(4)\times GL(1)$'s) in the upper-left and
lower-right, respectively.

The left-invariant (LI) current 
\begin{eqnarray}
(J^L_\mu)_{A}{}^{B} = (Z^{-1})_{A}{}^M\partial_\mu Z_M^{~B} = 
\begin{pmatrix}
({\bf J}_\mu)_a^{~b} & (j_\mu)_a^{~\bar{b}} \\
(\bar{j}_\mu)_{\bar{a}}^{~b} & (\bar{\bf J}_\mu)_{\bar{a}}^{~\bar{b}}
\end{pmatrix}\label{a1}
\end{eqnarray}
is decomposed into the coset part and the gauge potential part as
usual (For example, see \cite{AAR}). 
Before considering the decomposition, it is convenient to introduce the
decomposition of an arbitrary matrix $M_{ab}\in GL(4)$, 
\begin{eqnarray}
M_{ab} &=& - \frac{1}{4}\Omega_{ab}M^c_{~c} + M_{(ab)} + M_{\la ab\ra} 
\equiv - \frac{1}{4}\Omega{\rm tr}M + (M) + \la M \ra\,, 
\end{eqnarray}
into a trace part, symmetric part and traceless-antisymmetric part. 
By using this decomposition of $GL(4)$ part, 
the left-invariant current $(J^L)_A^{~B}$ is decomposed as follow: 
\begin{eqnarray}
&&\mbox{The coset part}: \quad ({\bf J}_\mu)^{\la ab \ra}\ ,~~
(\bar{\bf J}_\mu)^{\la \bar{a}\bar{b}\ra}\ ,~~ (j_\mu)^{a\bar{b}}~
,~~ (\bar{j}_\mu)^{\bar{a}b} \,, \\
&& \mbox{The gauge part}: \quad ({\bf J}_\mu)^{(ab)}~,~~ 
 (\bar{\bf J}_\mu)^{(\bar{a}\bar{b})}~~\cdots\mbox{$Sp(4)$'s}\,, 
\quad {\rm tr}{\bf J}~,~~
{\rm tr}\bar{\bf J}~~\cdots\mbox{$GL(1)$'s}\,.  \label{subgroup}
\end{eqnarray}

\par
\subsection{Constraints and Hamiltonian}

The Roiban-Siegel action of AdS-superstring is given 
by these LI currents as 
\bea
&& \hspace*{-1cm} 
S = \int\!\! d^2\sigma\,{\cal L}\,, \nnb\\
&& \hspace*{-1cm}
{\cal L} = \frac{1}{2}\left\{
-\sqrt{-g}g^{\mu\nu}({\bf J}^{\langle ab\rangle}_\mu {\bf J}_{\langle ab\rangle \nu}
-\bar{\bf J}^{\langle \bar{a}\bar{b}\rangle}_\mu \bar{\bf J}_{\nu~\langle \bar{a}\bar{b}\rangle })
 + \frac{k}{2}\epsilon^{\mu\nu}
 (E^{1/2}j^{ a\bar{b}}_\mu j_{\nu a\bar{b}~}
-E^{-1/2}\bar{j}^{\bar{a}{b}}_\mu \bar{j}_{\nu ~ \bar{a}{b}~})
\right\}\,, \label{stac}
\eea
with $k=\pm 1$ and $E={\rm sdet}Z_M^{~A}$\,.
We follow the notation in \cite{HK} where
canonical momentum for $Z_M^{~~A}$ is defined as
\bea
\Pi_A^{\ \ M}&=&\frac{\delta^r S}{\delta \dot{Z}_M^{\ \ A}}(-)^A\,, 
\label{ppp} 
\eea
and the canonical Poisson bracket is defined as
\bea
[ Z_M^{~~A}(\sigma),\Pi_{B}^{~~N}(\sigma')]_{\rm P}
=(-1)^{A}\delta_B^A\delta_M^N\delta(\sigma-\sigma')\,.\label{canpoi}
\eea
The Hamiltonian is given by
\bea
{\cal H}=\displaystyle\int\!\! d\sigma\, 
\left[\,\sum_{M,A} \Pi_A^{\ \ M}\dot{Z}_M^{\ \ A}(-)^A -{\cal L}\,\right]\,.\label{Hamil}
\eea

The supercovariant derivatives are
\bea
&&D_A^{~~B}=\Pi_A^{~~M}Z_M^{~~B}=
\left(\begin{array}{cc}
{\bf D}_a^{~~b}&  D_a^{~~\bar{b}}  \cr 
  \bar{D}_{\bar{a}}^{~~b} &  \bar{\bf D}_{\bar{a}}^{~~\bar{b}}
\end{array}\right)
\,,\label{covder0}
\eea
generating right (local) $GL(4|4)$ transformations
$\delta_{\Lambda(\sigma)} Z_M^{~~A}=[Z_M^{~~A},{\rm STr}D\Lambda(\sigma)]=
Z_M{}^{B}\Lambda_B{}^{A}(\sigma)$.
Its subgroups $GL(1),Sp(4),\overline{GL(1)},\overline{Sp(4)}$ 
are local gauge symmetries, and 
their generators are set to be constraints in our approach as follows: 
\bea
{\rm tr}{\bf D}=({\bf D})= {\rm tr}\bar{\bf D}=(\bar{\bf D})=0\,. 
\label{sp4gl1}
\eea
These correspond to gauge degrees of freedom \bref{subgroup}.

>From the definition of the canonical momenta \bref{ppp}
we have following primary constraints
\bea
\left\{\begin{array}{rcl}
F_a^{~~\bar{b}}&=&E^{1/4}D_a^{~~\bar{b}}+
 \frac{k}{2}E^{-1/4}(\bar{j}_{\sigma})^{\bar{b}}_{~a}=0\\
\bar{F}_{\bar{a}}^{~~{b}}&=&E^{-1/4}\bar{D}_{\bar{a}}^{~~{b}}+
\frac{k}{2} E^{1/4}({j}_{\sigma})^{b}_{~\bar{a}}=0
\end{array}\label{FFst}~~.\right.
\eea
First class part of the fermionic constraints are
\bea
\left\{\begin{array}{rcl}
B_a{}^{\bar{b}}&=&\left(\langle {\bf D}\rangle F+F\langle \bar{\bf D}\rangle\right)_a{}^{\bar{b}}
-\left(\langle \bar{\bf J}_{\sigma}\rangle \bar{F}+\bar{F}\langle {\bf J}_{\sigma}\rangle\right)^{\bar{b}}{}_a=0\\
\bar{B}_{\bar{a}}{}^{b}&=&\left(\langle \bar{\bf D}\rangle \bar{F}+\bar{F}\langle {\bf D}\rangle
\right)_{\bar{a}}{}^{b}
-\left(\langle {\bf J}_{\sigma}\rangle {F}+{F}\langle \bar{\bf J}_{\sigma}\rangle\right)^{b}{}_{\bar{a}}=0
\end{array}\label{FFst1st}~~\right.
\eea
generating the kappa-symmetry.
The Hamiltonian is given by
\bea
&&{\cal H}=\displaystyle\int\!\! d\sigma\,\left[
-\frac{2}{\sqrt{-g}g^{00}}A_{\perp}
-\frac{2g^{01}}{g^{00}}A_{\parallel}+{\rm tr}\{F\bar{\lambda}
+\bar{F}\lambda\}\right]\label{HamilSUST}\\
&&\left\{\begin{array}{ccl}
A_{\perp}&=&\frac{1}{2}{\rm tr}[\langle {\bf D}\rangle ^2 +\langle {\bf J}_{\sigma}\rangle ^2 
-\langle \bar{\bf D}\rangle ^2-\langle \bar{\bf J}_{\sigma}\rangle ^2 
]=0~~\label{AplSUST}\\
A_{\parallel}&=&{\rm tr}[\langle {\bf D}\rangle \langle {\bf J}_{\sigma}\rangle  
-\langle \bar{\bf D}\rangle \langle \bar{\bf J}_{\sigma}\rangle 
]=0~~\label{AppSUST}
\end{array}\right.
\eea
with multipliers $\lambda$'s determined consistently
\bea
\begin{array}{rcl}
\lambda&=&-\displaystyle\frac{2}{\sqrt{-g}g^{00}}(E^{-1/4}\bar{j}_{\sigma})
-\displaystyle\frac{2g^{01}}{g^{00}}(-E^{1/4}{j}_{\sigma})\\
\bar{\lambda}&=&-\displaystyle\frac{2}{\sqrt{-g}g^{00}}(-E^{1/4}j_{\sigma})
-\displaystyle\frac{2g^{01}}{g^{00}}(E^{-1/4}\bar{j}_{\sigma})
\end{array}\label{lmbd}~~~.
\eea
In the Hamiltonian \bref{HamilSUST}
 multipliers of the other first class constraints,
 the local $GL(1)$, $Sp(4)$, $\overline{GL(1)}$, $\overline{Sp(4)}$ constraints
in \bref{sp4gl1} and $B=\bar{B}=0$ in \bref{FFst1st}, are set to be zero.

Time development is determined by the Hamiltonian 
\bref{HamilSUST}, where we take the conformal gauge
$g_{00}+g_{11}=0=g_{01}$ as well as $E=1$
gauge,
\bea
{\cal H}=- \displaystyle\int\!\! d\sigma\, 
{\rm tr}\left[\frac{1}{2}\left\{\langle {\bf D}\rangle ^2 +\langle {\bf J}_{\sigma}\rangle ^2 
-\langle \bar{\bf D}\rangle ^2-\langle \bar{\bf J}_{\sigma}\rangle ^2 \right\}
+(\bar{D}\bar{j}_{\sigma}-Dj_{\sigma}+kj_{\sigma}\bar{j}_{\sigma})
\right]\,. 
\label{Hamiltonian}
\eea 
The equation of motion is given as
\bea
\dot{Z}_M{}^A=\left[Z_M{}^A,{\cal H}\right]=Z_M{}^B\Gamma_B{}^{A}\,,
\quad 
\Gamma_B{}^A=
\begin{pmatrix}\langle{\bf D}\rangle&-\bar{j}_{\sigma}\\
-j_{\sigma}&\langle \bar{\bf D}\rangle
\end{pmatrix}
\approx\begin{pmatrix}\langle{\bf D}\rangle&2kD\\
2k\bar{D}&\langle \bar{\bf D}\rangle
\end{pmatrix}\,,
\label{gamma}
\eea
where $\approx$ denotes that  the fermionic constraints \bref{FFst} are used.
\par

In order to construct explicitly an infinite set of conserved non-local
charges, we will consider left- and right-invariant currents below. 

\subsection{Left-invariant currents}

The left-invariant currents \bref{a1} in the conformal gauge 
are given as
\bea
J_\mu^{L}=Z^{-1}\partial_\mu Z=
\left\{\begin{array}{ccl}
(J_{\tau}^L)_A{}^B&=&\Gamma_A{}^B\\
(J_{\sigma}^L)_A{}^B&=&\left(Z^{-1}\partial_{\sigma} Z\right)_A{}^B
\end{array}\right.\label{a2}~~~
\eea
with $\Gamma_A{}^B$ in \bref{gamma}.
They satisfy the flatness condition by definition
\bea
\epsilon^{\mu\nu}\partial_\mu  J_\nu^{L}=
-\epsilon^{\mu\nu}Z^{-1}\partial_\mu Z Z^{-1} \partial_\nu Z
=-\epsilon^{\mu\nu}J^L_\mu J^L_\nu\,.
\eea
The equation of motion for the LI currents are
calculated by taking commutators with the Hamiltonian \bref{Hamiltonian}
\bea
&& \partial_{\tau} J_{\tau}^L = \nabla_{\sigma} J_{\sigma}^L+
2(2-k)\left[ \begin{pmatrix}
\langle {\bf D}\rangle&\\& \langle \bar{\bf D}\rangle
\end{pmatrix},
 \begin{pmatrix}
&D\\\bar{D}& 
\end{pmatrix}
\right] \nn \\
&& \hspace*{3cm}  -(2k-1)\left[ \begin{pmatrix}
\langle {\bf J}_{\sigma}\rangle&\\& \langle \bar{\bf J}_{\sigma}\rangle
\end{pmatrix},
 \begin{pmatrix}
&j_{\sigma}\\\bar{j}_{\sigma}& 
\end{pmatrix}
\right]~\label{eqLIc}
\eea
with the covariant derivatives on the LI currents: 
\bea
 \left\{\begin{array}{ccl}
\nabla_{\sigma} \langle{\bf J}_\mu\rangle&\equiv& 
\partial_{\sigma} \langle{\bf J}_\mu\rangle
+({\bf J}_{\sigma}) \langle{\bf J}_\mu\rangle
-\langle{\bf J}_\mu\rangle ({\bf J}_{\sigma})\\
 \nabla_{\sigma} j_\mu&\equiv& \partial_{\sigma} j_\mu 
 +({\bf J}_{\sigma})j_\mu-j_\mu (\bar{\bf J}_{\sigma})
-\frac{1}{4}({\rm Str} J_{\sigma})~j_\mu
\end{array}\right. \label{covder}
\eea
and similar expression holds for the barred sector.

\subsection{Right-invariant currents}

The right-invariant (RI) currents
are generators of the left multiplication on a coset element $Z_M{}^A$
which is the global GL(4${\mid}$4) symmetry 
$\delta_\Lambda Z_M{}^A=[Z_M{}^A,\displaystyle\int {\rm STr}~\Lambda
Z\Pi]_{\rm P}=\Lambda_M{}^N Z_N{}^A$.
Since the action \bref{stac}
 is written in terms of the LI currents,
which are manifestly invariant under the left multiplication,
the RI currents must be conserved.
The generator of the left multiplication is
the integration of the $\tau$-component of the RI currents 
\bea
(J^R_{\tau})_M{}^N=Z_M{}^A\Pi_A{}^N=Z_M{}^A D_A{}^B (Z^{-1})_B{}^N\,.
\label{b1}
\eea
The $\sigma$-component of the RI currents 
begins with  $(\partial_{\sigma} Z)Z^{-1}$
 which is neither invariant nor covariant under the right action on $Z_M{}^A$, 
 that is the local $[Sp(4)\times GL(1)]^2$\,.
Then we covariantize these local symmetries:  
\bea
(\partial_{\sigma} Z)Z^{-1} &\longrightarrow & (J^R_{\sigma})_M{}^N
\equiv ({\Delta}_{\sigma} Z)Z^{-1}
\equiv (\partial_{\sigma} Z) Z^{-1} +Z{\cal A}
Z^{-1}\nn \\
&& \hspace*{4.05cm} 
=Z_M{}^A(J^L_{\sigma}+{\cal A})_A{}^B (Z^{-1})_B{}^N\,, 
\label{Rsigma}
\eea
by introducing the  $[Sp(4)\otimes GL(1)]^2$-covariant derivative. 
The bosonic part of the gauge connection is 
obtained from the one on the LI currents \bref{covder} as follows: 
\bea
\left\{\begin{array}{ccl}
{\cal A}^{ab}&=&-({\bf J}_{\sigma})^{(ab)}+\frac{1}{4}\Omega^{ab}{\rm tr}{\bf J}_{\sigma}\\
{\cal A}^{\bar{a}\bar{b}}&=&-(\bar{\bf J}_{\sigma})^{(\bar{a}\bar{b})}
+\frac{1}{4}\Omega^{\bar{a}\bar{b}}{\rm tr}\bar{\bf J}_{\sigma}\\
\end{array}\right. \,. 
\label{aa1} \eea Then the $\sigma$-component of the covariant RI current
is invariant under the local symmetry transformation $Z\to Zh^{-1}$ with
$h\in$$[Sp(4)\otimes GL(1)]^2$ by the gauge field transformation as
${\cal A}+(\partial_\mu h^{-1})h~\to~{\cal A}$.  Bosonic part of the
gauge connection ${\cal A}$ is an element of the stability group.

The current only with \bref{aa1} is however not conserved.
For a superstring the existence of the fermionic constraints
requires the non-zero fermionic elements of ${\cal A}$.
This fact is related to that the super-coset space is not a symmetric space. 
We determine the fermionic components of the gauge connection ${\cal A}$
from the current conservation law, 
\begin{eqnarray}
\partial^\mu J_\mu^R = - \partial_{\tau} J_{\tau}^R 
+ \partial_{\sigma}J_{\sigma}^{R} = 0\,. 
\label{clr}
\end{eqnarray}
The first term in (\ref{clr}) is 
\begin{eqnarray}
&&\partial_{\tau} J^R_{\tau} = Z
\Biggl\{ \partial_{\sigma} \begin{pmatrix} \langle {\bf
J}_{\sigma}\rangle&\frac{k}{2}j_{\sigma}\\
\frac{k}{2}\bar{j}_{\sigma}&\langle\bar{\bf J}_{\sigma}\rangle
\end{pmatrix}
+\left[~J_{\sigma}^L,  \begin{pmatrix}
\langle {\bf J}_{\sigma}\rangle&\frac{k}{2}j_{\sigma}\\
\frac{k}{2}\bar{j}_{\sigma}&\langle\bar{\bf J}_{\sigma}\rangle
\end{pmatrix}
\right] \nn \\
&& \hspace*{4cm} +2(k-1)\left[~
 \begin{pmatrix}\langle {\bf D}\rangle&\\&\langle \bar{\bf D}\rangle
\end{pmatrix}\,,
 \begin{pmatrix}&D\\ \bar{D}&\end{pmatrix}
\right] \Biggr\} Z^{-1}\,.  
\end{eqnarray}
The second term in (\ref{clr}) is 
\begin{eqnarray}
\partial_{\sigma} J^R_{\sigma}=Z \left\{
\partial_{\sigma}(J^L_{\sigma}+{\cal A})+\left[J^L_{\sigma},{\cal
A}\right] \right\}Z^{-1}\,.  \end{eqnarray}
Then we can read off 
\begin{eqnarray}
J^L_{\sigma}+{\cal A}=\begin{pmatrix} \langle {\bf
J}_{\sigma}\rangle&\frac{k}{2}j_{\sigma}\\
\frac{k}{2}\bar{j}_{\sigma}&\langle\bar{\bf J}_{\sigma}\rangle
\end{pmatrix}\equiv \langle J^L_{\sigma}\rangle\,,
\end{eqnarray}
and the $\kappa$-parameter $k=1$ is chosen. 
The fermionic component of the connection is determined as 
\bea
\left\{\begin{array}{ccl}
{\cal A}^{a\bar{b}}&=&(\frac{k}{2}-1)j^{a\bar{b}}_{\sigma}\\
{\cal A}^{\bar{a}b}&=&(\frac{k}{2}-1)\bar{j}^{\bar{a}b}_{\sigma}\\
\end{array}\right.\,.\label{aa2}
\eea

Next we examine the flatness for the RI currents.
By using the equation of motions 
\bea
\partial_{\tau} \langle J^L_{\sigma}\rangle
=\partial_{\sigma} D+[D,{\cal A}]-\left[
\begin{pmatrix}&D\\\bar{D}&\end{pmatrix}
,\langle J^L_{\sigma}\rangle
\right]\,,
\eea
the flatness is given as
\bea
\partial_{\tau} J^R_{\sigma}-\partial_{\sigma} J^R_{\tau}&=&
Z\left\{\partial_{\tau} \langle J^L_{\sigma}\rangle
+\left[\Gamma, \langle J^L_{\sigma}\rangle
\right]-\partial_{\sigma} D-\left[J_{\sigma}^L, D\right]
\right\}Z^{-1}\nn\\
&=&2Z\left\{
\left[ D,\langle J^L_{\sigma}\rangle\right]
\right\}Z^{-1}\nn\\
&=&2(J_{\tau}^RJ_{\sigma}^R-J_{\sigma}^RJ_{\tau}^R)~~~.\label{RIcflat}
\eea
The front factor $2$ can be absorbed
by rescaling $J^R \to (1/2)J^R$ reproducing
the usual flatness condition $dJ^R=J^RJ^R$. 

In the next section, we will construct an infinite set of conserved
non-local charges by using the RI currents.

\section{Super Yangian and Classical Integrability of AdS-String} 

In this section we will discuss super Yangian algebra of type IIB
superstring on the AdS$_5\times$ S$^5$\,.
We will construct an infinite set of non-local charges in the canonical
formalism, and calculate the Poisson bracket algebra of them. 
The resulting algebra is a Hopf-Poisson algebra called the Yangian
algebra. We can show the Yangian algebra in terms of the supermatrix,
i.e., the super Yangian algebra of the AdS-superstring. 
 To begin with, we explicitly
construct non-local charges. Then the Yangian algebra is calculated by
using the Poisson bracket. The Serre relation is also satisfied.  In
addition, the Lax pair and the transfer matrix are presented, the
existence of them implies a classical integrability of AdS-superstring.

\subsection{Construction of Non-Local Charges}

Here we shall construct non-local charges following 
Brezin, Izykson , Zinn-Justin and Zuber \cite{BIZZ}.
In previous section we have obtained the conserved and flat current
which is the RI currents
\bea
(J^R)_\mu=\left\{\begin{array}{ccl}
J^R_{\tau}&=&Z\Pi\\
J^R_{\sigma}&=&(\partial_{\sigma} Z) Z^{-1} +Z{\cal A} Z^{-1}=Z(J_\sigma^L+{\cal A})Z^{-1}
\end{array}\right.~~\label{31}
\eea
with
\bea~~
J_\sigma^L+{\cal A}
=\begin{pmatrix}
\langle {\bf J}_{\sigma}\rangle&\frac{1}{2}j_{\sigma}\\
\frac{1}{2}\bar{j}_{\sigma}&\langle\bar{\bf J}_{\sigma}\rangle
\end{pmatrix}\,.
\eea
By defining the covariant derivative
\bea
{\cal D}_\mu=\partial_\mu-2J_\mu^R\,, 
\eea
the flatness condition \bref{RIcflat}
and the conservation law are expressed as
\bea
\epsilon^{\mu\nu}[{\cal D}_\mu,{\cal D}_\nu]=0\,, \quad 
[\partial^\mu,{\cal D}_\mu]=0\,.
\label{ddd1}
\eea
Conserved non-local currents are constructed as follows:
Beginning with a conserved current 
\bea
({\cal J}_n)_\mu=\epsilon_{\mu\nu}\partial^\nu\chi_n  \quad 
(n\geq 0)\,, \label{potchi}
\eea
another conserved current can be constructed as
\bea
({\cal J}_{n+1})_\mu={\cal D}_\mu \chi_n\,,
\eea
where equations in \bref{ddd1} are used.
We take $({\cal J}_{-1})_\mu=0$ and $\chi_{-1}=-\frac{1}{2}$\,, then
the conserved non-local currents are given as
\bea
\left\{\begin{array}{rcl}
({\cal J}_0)_\mu(\sigma)&=&J^R_\mu(\sigma)\\
({\cal J}_1)_\mu(\sigma)&=&-\epsilon_{\mu\nu}(J^R)^\nu(\sigma)-
2J^R_\mu(\sigma)\displaystyle\int_{}^{\sigma}\!\!d\sigma'\, 
(J^R)^\tau(\sigma')\\
({\cal J}_2)_{\mu}(\sigma) &=& - J_{\mu}^R - 2 \epsilon_{\mu\nu}
 J^{R\nu} 
\displaystyle\int^{\sigma}\!\!d\sigma'\,(J^R)^{\tau}(\sigma') 
+ 2 J^R_{\mu}\displaystyle\int^{\sigma}\!\! d\sigma'\, 
(J^R)_{\sigma}(\sigma') \\
&& \qquad \qquad 
+ 4 J_{\mu}^R
\displaystyle\int^{\sigma}\!\!d\sigma'\,(J^R)^{\tau}(\sigma')
\displaystyle\int\!\!
d\sigma''\,(J^R)^{\tau}(\sigma'')
\\
\vdots \end{array}\right.\,. \label{37}
 \eea 
There exists infinite number of the conserved non-local charges $Q_n$
which are the $\sigma$ integration of the $\tau$-component of the 
non-local currents, $\int\!d\sigma ({\cal J}_n)_\tau$\,.

It is commented about the range of the spatial direction we
consider. In studies of Yangian of sigma models such as chiral
principle models and $O(N)$ sigma models, the integral for the spatial
direction is performed from $-\infty$ to $+\infty$\,. 
But we 
would like to consider a closed string satisfying a periodic boundary
condition\footnote{ 
The classical Poisson bracket may be inconsistent
with non-trivial boundary conditions when the current algebra includes 
non-ultra local terms \cite{dVEM:CMP}.  
As an example, the boundary condition $ X_a(x)\mid_{\pm\infty}={\rm
const.}  $ i.e., $\partial_x X_a(x)\mid_{\pm\infty}=0$\,, is
incompatible with the definition of the Poisson bracket at the boundary
as follows:
\begin{eqnarray}
[X_a(x),P^b(y)]_{\rm P}=\delta_a^b \delta (x-y)~~~\to~~ [\partial_x
X_a(x),P^b(y)]_{\rm P}\mid_{x={\pm\infty}}=0 \neq \delta_a^b
\partial_x\delta (x-y)\mid_{x={\pm\infty}}\,. 
\end{eqnarray}
The modification of the
Poisson bracket implies to change the vacua in the theory
\cite{Schwinger}. This view is proposed in \cite{FR}.}  
as 
\begin{eqnarray}
Z_M{}^A(\tau,\sigma)=Z_M{}^A(\tau,\sigma+2\pi) \label{zzz}
\end{eqnarray} 
with $-\infty\leq
\tau \leq \infty$ and $0\leq \sigma\leq 2\pi$
which are the coordinates of a cylinder. 
We naively take this condition\footnote{
In the case of the finite interval $[0,2\pi]$, we possibly should modify 
the $J^R$, for example, as follows: 
\bea
(J^R)_\sigma(\tau,\sigma)=\partial_\tau \chi_0(\tau,\sigma)\,, \quad 
\chi_0(\tau,\sigma)=\displaystyle\int^\sigma_0\!\!d\sigma'\,(J^R)_\tau(\tau,\sigma')
+\displaystyle\int_{-\infty}^{\tau}\!\!\! d\tau'\,(J^R)_\sigma(\tau',\sigma=0)\,.\label{chi0chi0}
\eea 
The non-local charge $Q_1$ constructed from the above currents 
may be conserved and satisfy \bref{318}.
However the cubic Serre relation of the Yangian and \bref{q1q1} 
may contain additional terms and so we need some efforts to justify 
\bref{chi0chi0}. 
In this paper we present the computation without the extra terms 
in \bref{chi0chi0} as the first step. }, but then 
the nonabelian generators may be broken by finite-size effects.

Hence the super Yangian generators\footnote{Non-local charges and local
charges in supersymmetric NLSM$_2$ are discussed in \cite{SNLSM, EHMM}.
But the supersymmetries are not space-time but world-sheet.} are given
by
\begin{eqnarray}
 Q_{0} &\equiv& {\displaystyle\int_0^{2\pi}}\!\!\!\! d\sigma\, 
({\cal J}_0)_\tau 
={\displaystyle\int_0^{2\pi}}\!\!\!\!
d\sigma\, J^R_\tau(\sigma)\,, \\ 
 Q_{1} &\equiv&- {\displaystyle\int_0^{2\pi}}\!\!\!\! 
d\sigma\,({\cal J}_1)_\tau 
= {\displaystyle\int_0^{2\pi}}\!\!\!\!  
d\sigma\,J^R_\sigma (\sigma) -2{\displaystyle\int_0^{2\pi}}\!\!\!\! 
d\sigma \displaystyle\int_0^{\sigma}  d\sigma'\, 
\left[J^R_\tau(\sigma),\, J^R_\tau(\sigma')\right]\nn\\
&=&{\displaystyle\int_0^{2\pi}}\!\!\!\! d\sigma\,J^R_\sigma (\sigma) 
-{\displaystyle\int_0^{2\pi}}\!\!\!\! 
d\sigma \displaystyle\int_0^{2\pi}\!\!\!\!  d\sigma'\,
\epsilon(\sigma - \sigma')
\left[J^R_\tau(\sigma),\, J^R_\tau(\sigma')\right]\,, 
\end{eqnarray}
where we have introduced the function $\epsilon(x) = \th(x) -\th(-x)$\,, 
so that 
\begin{eqnarray}
\epsilon(x) = 1 \quad (x > 0)\,, \qquad 
 \epsilon(-x) = -1 \quad (x < 0)\,. 
\end{eqnarray}

We will calculate the Poisson bracket algebra of the charges $Q_0$ and
$Q_1$ below.  
The $\kappa$-invariance as well as the gauge invariances
are confirmed explicitly by the Poisson bracket between
charges and the gauge generators which are left-invariant currents.
Firstly, we will consider the bosonic case to recollect
the (bosonic) Yangian algebra.  After the discussion of the bosonic
case, we will consider the supersymmetric case by taking care of the
gradation of fermionic variables.

\subsection{Bosonic Yangian Algebra}

Before going to the full supersymmetric case, let us consider a bosonic
part, which is described like a principle chiral model, 
contained in the Roiban-Siegel AdS-superstring. 
For simplicity, we will concentrate on the upper left part (AdS$_5$). 

The Poisson bracket of a bosonic (upper-left) part is given by  
\begin{eqnarray}
\left[{\bf z}_{m}^{~a}(\sigma),\, {\bf \pi}_{b}^{~n}(\sigma')\right]_{\rm P} = 
\del_{m}^{n}\del^{a}_{b}\del(\sigma - \sigma')\,,
\end{eqnarray}
where the ``delta function'' has a periodicity with $2\pi$\,. 
The bosonic RI currents are 
\bea
(J^R)_\mu=\left\{\begin{array}{ccl}
(J^R_{\tau})_m{}^n&=&{\bf z}_m{}^a{\bf \pi}_a{}^n\\
(J^R_{\sigma})_m{}^n&=&(\partial_{\sigma} {\bf z})_m{}^a( {\bf z}^{-1})_a{}^n 
+{\bf z}_m{}^a{\cal A}_a{}^b ({\bf z}^{-1})_b{}^n
=({\bf z})_m{}^a({\bf J}_\sigma^L+{\cal A})_a{}^b({\bf z}^{-1})_b{}^n
\end{array}\right.~~\label{311}
\eea
with ${\bf z}_m{}^a({\bf z}^{-1})_a{}^n=\delta_m^n$. 
In the calculation of Yangian algebra, it is convenient to use the 
Poisson bracket in terms of ${J^R_\tau}_{m}^{~n}$ and ${J^R_\sigma}_{m}^{~n}$: 
\begin{eqnarray}
 \left[\int({J^R_\tau})_m^{~n}(\sigma),\, ({J^R_\tau})_l^{~k}(\sigma')\right]_{\rm P} 
&=& \left(\del_m^{k}({J^R_\tau})_l^{~n}
- \del_l^{n}({J^R_\tau})_m^{~k}
 \right)(\sigma'), \\ 
 \left[\int
 ({J^R_\tau})_m^{~n}(\sigma),\, ({J^R_\sigma})_l^{~k}(\sigma')\right]_{\rm P}& = &
\left(\del_m^k ({J^R_\sigma})_l^{~n}
 - \del_l^n ({J^R_\sigma})_m^{~k}
 \right)(\sigma')\nn\\
 \left[({J^R_\sigma})_m^{~n}(\sigma),\, ({J^R_\sigma})_l^{~k}(\sigma')\right]_{\rm P}& = &
0\nn
\end{eqnarray}
It is denoted that $\left[\int J^R_\mu, {\cal A}\right]=0$
because the gauge connection ${}\cal A$ is written in terms of 
the LI currents, $[\int (J^R_{\mu})_m^{~n},\,(J_{\nu}^L)_a^{~b}] =
0$\,. As a remark, we have to be careful to treat the non-ultra local
term in the Poisson bracket of $J_{\tau}^R$ and $J_{\sigma}^R$\,, as in
the principle chiral models. For the treatment for the non-ultra local
term, see Appendix \ref{app:reg}. 

We obtain the Yangian algebra:
\begin{eqnarray}
&& \left[Q_{0m}^{~~n},\, Q_{0l}^{~~k}\right]_{\rm P} = \delta_m^{k}
Q_{0l}^{~~n} - \delta_l^{n} Q_{0m}^{~~k}\,, \\ 
&& \left[Q_{0m}^{~~n},\, Q_{1l}^{~~k}\right]_{\rm P} = \delta_m^{k}
Q_{1l}^{~~n} - \delta_l^{n}Q_{1m}^{~~k}\,,\label{318} \\ 
&& \left[Q_{1m}^{~~n},\, Q_{1l}^{~~k}\right]_{\rm P} = 
\delta_m^{k} Q_{2l}^{~~n} - \delta_l^{n} Q_{2m}^{~~k} 
+ 4(Q_{0m}^{~~k}Q_{0l}^{~~p}Q_{0p}^{~~n} 
- Q_{0l}^{~~n}Q_{0m}^{~~p}Q_{0p}^{~~k})\,,  \label{q1q1}
\end{eqnarray}
where we have defined a tri-local charge $Q_2$ as 
\begin{eqnarray}
&& Q_{2m}^{~~k}(\tau) \equiv 
-4 Q_{0m}^{~~k} +2 {\displaystyle\int_0^{2\pi}} 
d\sigma{\displaystyle\int_0^{2\pi}} 
d\sigma'\,\epsilon(\sigma-\sigma')\left[J^R_\tau(\sigma),\, J^R_\sigma(\sigma')
\right]_{m}^{~k} \label{Q2} \\ 
&& \qquad \qquad
+ 4{\displaystyle\int_0^{2\pi}} d\sigma
{\displaystyle\int_0^{2\pi}} d\sigma'
{\displaystyle\int_0^{2\pi}} d\sigma''\, \epsilon(\sigma - \sigma')
\epsilon(\sigma' - \sigma'') ({J^R_\tau})_{m}^{~p}(\sigma)({J^R_\tau})_p^{~p'}(\sigma')
({J^R_\tau})_{p'}^{~k}(\sigma'')\,. \nn 
\end{eqnarray}
In the calculation of the algebra, it is convenient to use 
the identity:
\begin{eqnarray}
\epsilon(\sigma'-\sigma)\epsilon(\sigma - \sigma'') 
+ \epsilon(\sigma - \sigma')\epsilon(\sigma' - \sigma'') 
+ \epsilon(\sigma  - \sigma'')\epsilon(\sigma'' - \sigma') = -1\,. 
\end{eqnarray}

The Serre relation, which is needed to ensure the integrable structure 
is the standard Yangian algebra, is shown, up to the ambiguity of linear
term with respect to $Q_0$, as follows:
\begin{eqnarray}
&& [Q_{0m}^{~~n},\,[Q_{1l}^{~~k},\,Q_{1p}^{~~q}]_{\rm P}]_{\rm P}
- [Q_{1m}^{~~n},\,[Q_{0l}^{~~k},\, Q_{1p}^{~~q}]_{\rm P}]_{\rm P} \nn \\ 
&=&4 [Q_{0m}^{~~n},\,[Q_{0l}^{~p'}Q_{0p'}^{~~k},\,Q_{0p}^{~~p''}
Q_{0p''}^{~~q}]_{\rm P}]_{\rm P} 
- 4[Q_{0m}^{~~p'}Q_{0p'}^{~~n},\,[Q_{0l}^{~~k},\,Q_{0p}^{~~p''}
Q_{0p''}^{~~q}]_{\rm P}]_{\rm P}\,.   
\end{eqnarray}
The point is that the higher order charge $Q_2$, which appears
in the $[Q_1,Q_1]$, disappears in the above Serre relation.              
The fifth linear term in $Q_0$ becomes the tri-linear in $Q_0$ \footnote{ 
It is shown  in the adjoint index of the matrix as
\bea
[Q_1^a,[Q_1^b,Q_0^c]]+{\rm cyclic ~in} ~a,b,c
&=&12 f^{arl}f^{bsk}f^{ctm}f^{trs}Q_0^lQ_0^mQ_0^k
\eea
where $f^{abc}$ with $[Q_0^a,Q_0^b]=f^{abc}Q_0^c$ is the totally antisymmetric structure constant 
for the $SL(4)$ group after the  $GL(1)$ gauge fixed.
}. 

In the next subsection we will consider the full supermatrix case by
following the above-mentioned regularization and order of limits.

\subsection{Super Yangian Algebra}

Let now us calculate the Poisson bracket algebra of non-local charges 
in the supersymmetric case, and show the super Yangian algebra. 

Using the Poisson bracket \bref{canpoi}, 
the current algebra in terms of the supermatrix is given by 
\begin{eqnarray}
&&\hspace{-0.5cm}
[\int(J^R_\tau)_M^{~N}(\sigma),\, (J^R_\tau)_L^{~K}(\sigma')\}_{\rm P} 
= (-)^N\left[(-)^{(N+L)(1+M+L)}\del_M^{K}(J^R_\tau)_L^{~N} 
- \del_L^{N}(J^R_\tau)_M^{~K}
\right](\sigma') ,\nn \\ 
&& \hspace{-0.5cm} 
[\int (J^R_\tau)_M^{~N}(\sigma),\, ({J^R_\sigma})_L^{~K}(\sigma')\}_{\rm P} = 
(-)^N\left[(-)^{(N+L)(1+M+L)}
\del_M^K {J^R_\sigma}_L^{~N} 
 - \del_L^N {J^R_\sigma}_M^{~K} 
\right](\sigma') 
. 
\end{eqnarray}
For the super case the fermionic constraints exist \bref{FFst}.  
Half of
it are first-class constraints generating the $\kappa$-symmetry \cite{HK} as 
the same situation of the flat  superstring,
and the non-local charges commute with it since the fermionic constraints 
are LI currents,
$\{\int J^R_\mu,J^L_\nu\}=0$. 
Another half constraints
are second-class constraints which requires to modify the Poisson
bracket to the Dirac bracket: $\{,\}_{\rm P}\to \{,\}_{\rm Dirac}$.
Fortunately the Dirac brackets of the RI currents are equal to the
Poisson bracket, $\{\int J^R_\mu,*\}_{\rm Dirac}=\{\int J^R_\mu,*\}_{\rm P}$\,,
since the fermionic second-class constraints are also 
LI currents. 
Non-ultra local terms also appears in the
supersymmetric case, but the prescription is the same as in the bosonic
case.  

The super Yangian algebra i.e., the Poisson bracket algebra of non-local
charges in the case of super matrix, is given by 
\begin{eqnarray}
&& [Q_{0M}^{~~~N},\,Q_{0L}^{~~K} \}_{\rm P} = (-)^N \left[(-)^{(N+L)(1+M+L)}
\delta_M^{K}Q_{0L}^{~~N} - \delta_L^{N}Q_{0M}^{~~~K} \right]\,, \\ 
&& [Q_{0M}^{~~~N},\,Q_{1L}^{~~K}\}_{\rm P} = (-)^N\left[(-)^{(N+L)(1+M+L)}
\delta_M^{K}Q_{1L}^{~~N} - \delta_L^{N}Q_{1M}^{~~~K} \right]\,, \\ 
&& [Q_{1M}^{~~~N},\,Q_{1L}^{~~K}\}_{\rm P} = (-)^N 
[(-)^{(N+L)(1+M+L)}\delta_M^{K} Q_{2L}^{~~N} -
\delta_{L}^{N} Q_{2M}^{~~~K}] \\ && 
\hspace*{3cm} + (-)^{N}(-)^{(N+L)(1+M+L)}4
[Q_{0L}^{~~P}Q_{0P}^{~~N}Q_{0M}^{~~K} 
- Q_{0L}^{~~N}Q_{0M}^{~~P}Q_{0P}^{~~K}]\,, \nn
\end{eqnarray}
where we have defined $Q_2$ as 
\begin{eqnarray}
&& Q_{2M}^{~~~K}(\tau) \equiv 
-4 Q_{0M}^{~~~K} + 2{\displaystyle\int_0^{2\pi}} 
d\sigma{\displaystyle\int_0^{2\pi}}  
d\sigma'\,\epsilon(\sigma-\sigma')[(J^R_\tau)
(\sigma),\, (J^R_\sigma)(\sigma')]_{M}^{~~K} 
 \\ && \qquad \quad \quad 
+4 {\displaystyle\int_0^{2\pi}} d\sigma
{\displaystyle\int_0^{2\pi}} d\sigma'
{\displaystyle\int_0^{2\pi}} d\sigma''\, \epsilon(\sigma - \sigma')
\epsilon(\sigma' - \sigma'') (J^R_\tau)_M^{~~P}(\sigma)(J^R_\tau)_P^{~~P'}(\sigma')
(J^R_\tau)_{P'}^{~~K}(\sigma'')\,. \nn
\end{eqnarray}
When we introduce the notation, following \cite{BR},   
\begin{eqnarray}
\widehat{Q}_{iM}^{~~~N} \equiv (-)^M Q_{iM}^{~~~N} \quad 
(i=0,1,2,\ldots,)\,, 
\end{eqnarray}
the super Yangian algebra is rewritten as follows: 
\begin{eqnarray}
&& [\widehat{Q}_{0M}^{~~~N},\,\widehat{Q}_{0L}^{~~K} \}_{\rm P} 
= (-)^{(M+N)(L+K)}
\delta_M^{K}\widehat{Q}_{0L}^{~~N} - \delta_L^{N}\widehat{Q}_{0M}^{~~~K}\,, \\ 
&& [\widehat{Q}_{0M}^{~~~N},\,\widehat{Q}_{1L}^{~~K}\}_{\rm P} 
= (-)^{(M+N)(L+K)}
\delta_M^{K}\widehat{Q}_{1L}^{~~N} - \delta_L^{N}\widehat{Q}_{1M}^{~~~K}\,, \\ 
&& [\widehat{Q}_{1M}^{~~~N},\,\widehat{Q}_{1L}^{~~K}\}_{\rm P} = 
(-)^{(M+N)(L+K)}\delta_M^{K} \widehat{Q}_{2L}^{~~N} -
\delta_{L}^{N} \widehat{Q}_{2M}^{~~~K} \\ && 
\hspace*{3.3cm} + (-)^{(M+N)(L+K)}4
[\widehat{Q}_{0L}^{~~P}\widehat{Q}_{0P}^{~~~N}\widehat{Q}_{0M}^{~~~K} 
- \widehat{Q}_{0L}^{~~N}\widehat{Q}_{0M}^{~~~P}\widehat{Q}_{0P}^{~~~K}]\,. \nn
\end{eqnarray}
This form is the familiar expression of the $GL(4|4)$
\footnote{
The global symmetry of the AdS superstring is $PSL(4|4)$ as discussed 
by Metsaev and Tseytlin \cite{MT},
and in the Roiban-Siegel formalism auxiliary degrees of freedom for two $GL(1)$'s
are introduced for the simpler linear realization of $GL(4|4)$.
The $GL(1)$ symmetry is anomalous for the AdS superstring and the D-branes
classically by the WZ term \cite{HKAdSD}.
The superparticle system, 
which governs the massless sectors
of the ${\cal N}=4$ Yang-Mills system,
 contains 
manifest $GL(4|4)$ symmetry including $GL(1)$'s in the Roiban-Siegel formalism.
The $GL(1)$ may be broken in the quantum level, since the  measure is not invariant 
under it.
 We are grateful to Nathan Berkovits and Warren Siegel for fruitful discussions
 on this point.       } 
super Yangian
algebra (often denoted as $Y(gl(4|4))$ or simply $Y(4|4)$).  The super
Yangian for $GL(M|N)$ was first defined by Nazarov \cite{Nazarov}.  It
can be obtained as the generalization of the construction for the
Yangian $Y(gl(M))$ (bosonic case), based on the Lie algebra $gl(M)$\,,
to the case of the super Lie algebra $gl(M|N)$\,. Its representations
are studied by Zhang \cite{Zhang}.

In addition, we can show the super Serre relation in the full
supermatrix case:
\begin{eqnarray}
&& [Q_{0M}^{~~~N},\,[Q_{1L}^{~~~K},\,Q_{1P}^{~~~Q}\}_{\rm P}\}_{\rm P}
- [Q_{1M}^{~~~N},\,[Q_{0L}^{~~~K},\, Q_{1P}^{~~~Q}\}_{\rm P}\}_{\rm P}  \\ 
&=&  [Q_{0M}^{~~~N},\,[Q_{0L}^{~~~P'}Q_{0P'}^{~~~K},\,Q_{0P}^{~~~P''}
Q_{0P''}^{~~~Q}\}_{\rm P}\}_{\rm P} 
- [Q_{0M}^{~~~P'}Q_{0P'}^{~~~N},\,[Q_{0L}^{~~K},\,Q_{0P}^{~~~P''}
Q_{0P''}^{~~~Q}\}_{\rm P}\}_{\rm P}\,.   \nn 
\end{eqnarray}
This is a super extension of the standard Serre relation in the bosonic
case. It should be noted that the Serre relation is not modified up to
the replacement of the commutator while the index factors appear in the
case of super extension of Jacobi identity.  
This fact implies that the
structure of Hopf algebra is preserved under the super extension. As a
final remark, the above Serre relation is not modified even if we
replace $Q_{iM}^{~~N}$ by $\widehat{Q}_{iM}^{~~N}$\,.

\subsection{Lax Pair and Transfer Matrix}

Here, we shall consider the integrable structure of AdS-superstring.  

An alternative to the iterative definition of the non-local charges is
to utilize the Lax pair given by 
\begin{eqnarray}
&& (L_{\tau})_{ M}^{~~~N}(\sigma ; \lambda) \equiv 
\frac{\lambda}{\lambda^2 -1}
\left[\lambda (J^R_\tau)_{M}^{~~N} + (J^R_\sigma)_{M}^{~~N}\right]\,, \\ 
&& (L_{\sigma})_{ M}^{~~~N}(\sigma ; \lambda) 
\equiv \frac{\lambda}{\lambda^2 -1}
\left[\lambda (J^R_\sigma)_{M}^{~~N} +(J^R_\tau)_{M}^{~~N}\right]\,. 
\end{eqnarray}
The zero-curvature condition and the conservation law are equivalent to 
the condition: 
\begin{eqnarray}
\left[\partial_\tau + L_\tau,\, \partial_\sigma + L_\sigma\right] = 0\,.
\end{eqnarray}
By using the Lax pair, a transfer matrix $T(\sigma,\sigma';\lambda)$ 
can be defined as a solution of the equation 
\begin{eqnarray}
\left[\partial_\sigma + L_\sigma(\sigma;\lambda)\right]T(\sigma,\sigma';\lambda) = 0\,, \quad 
T(\sigma,\sigma;\lambda) ={\bf 1}\,. 
\label{eq-trans}
\end{eqnarray}
The solution of (\ref{eq-trans}) is 
\begin{eqnarray}
T(\sigma,\sigma';\lambda) = {\rm P}\exp\left(-\int^\sigma_{\sigma'} \!\! d\xi\, 
L_\sigma(\xi;\lambda)\right)\,,
\end{eqnarray}
where the symbol P denotes the equal-time path ordering in terms of 
$\epsilon(\sigma)$\,.  
We can re-derive non-local charges by expanding the following relation: 
\begin{eqnarray}
T(\lambda) \equiv T(2\pi,0;\lambda) =
 \exp\left(\sum_{n=0}^{\infty}\lambda^{n+1}Q_n 
\right)\,. 
\end{eqnarray} 
The $Q_0$ and $Q_1$ are obtained from the first and second order parts
with respect to $\lambda$\,, respectively.
The gauge invariance of all non-local charges follows the fact that
$T(2\pi,0;\lambda)$ commute with the gauge generators.

It is possible to derive a classical $r$-matrix of the AdS-superstring by
calculating the Poisson bracket of transfer matrices.  
Though we are confronted with the subtlety as in the case of principle
chiral models \footnote{If there is no non-ultra-local term, for example
in the case of generalized Gross-Neveu model, then we do not encounter
this subtlety \cite{dVEM:PLB}.  The classical $r$-matrices are well-defined
\cite{dVEM:PLB} and the classical Yang-Baxter equation is also satisfied
\cite{dVEM:NPB}. } \cite{dVEM:CMP} because of the non-ultra local terms,
we can avoid it by following the prescription given in
\cite{FR,Maillet,DNN}.  Indeed, after small algebra, 
we can readily show the following Poisson bracket for the 
$L_{\sigma}(\sigma;\lambda)$: 
\begin{eqnarray}
&& [\int\!\!
(L_{\sigma})_M^{~~N}(\sigma;\lambda),\,(L_{\sigma})_L^{~~K}
(\sigma';\mu)\}_{\rm P} \nn  \\
&=& \frac{\lambda\mu}{\lambda-\mu}(-)^N
\Biggl[ \frac{1}{\mu^2-1}\Bigl\{
(-)^{(N+L)(1+M+L)}\delta_M^K (L_{\sigma})_L^{~~N} (\sigma';\lambda) -
\delta_L^N(L_{\sigma})_M^{~~K}(\sigma';\lambda)\Bigr\} \\
&& \qquad \qquad \qquad -
\frac{1}{\lambda^2-1}\Bigl\{ (-)^{(N+L)(1+M+L)}\delta_M^K
(L_{\sigma})_L^{~~N}(\sigma';\mu) - \delta_L^N
(L_{\sigma})_M^{~~K}(\sigma';\mu)\Bigr\}\Biggr]\,. \nn 
\end{eqnarray}
Following the paper \cite{DNN}\,, we can read off the r-matrix defined by 
\begin{eqnarray}
 r(\lambda,\mu)_{ML}^{NK} \equiv \frac{\lambda\mu}{\lambda-\mu}\delta_L^N\delta_M^K(-)^{NK} \,. 
\end{eqnarray} 
This $r$-matrix is well defined and obey 
the classical Yang-Baxter equation, 
\begin{eqnarray}
r(\lambda,\mu)_{MQ}^{NP}~r(\mu,\nu)_{QL}^{PK} 
+ r(\mu,\nu)_{MQ}^{NP}~r(\nu,\lambda)_{QL}^{PK}
 + r(\nu,\lambda)_{MQ}^{NP}~r(\lambda,\mu)_{QL}^{PK} =0\,,  
\end{eqnarray} 
as it is ensured by the structure of $Y(4|4)$\,. Thus the classical
integrability of the AdS-superstring has been shown.

\section{Conclusion and Discussion} 

We have discussed a classical integrability of the type IIB superstring
theory on the AdS$_5\times$ S$^5$ background from the viewpoint of Yangian
symmetry. The Roiban-Siegel formulation has been used to explicitly
construct an infinite set of non-local charges, the first two of which
are the generators of the Yangian algebra. 
We have constructed explicitly non-local charges by carefully treating
the Wess-Zumino term and the constraint conditions intrinsic to the
AdS-superstring.  
The obtained non-local charges are $\kappa$-invariant as well as the 
gauge invariant. 
The Hamiltonian \bref{Hamiltonian}
 also commutes with these non-local charges.

The existence of the fermionic constraints requires the
local gauge group to be super-group as derived in
\bref{aa1} and \bref{aa2}.  Since the direct origin of the fermionic
components of gauge connection ${\cal A}$ is the fermionic constraints,
this local super-group may be related to the local constraints algebra
which is shown to exist in the AdS-superstring \cite{HK} as well as the
flat superstring.  However this local gauge group is the stability group
of the coset at least for the bosonic part, so this super-group is a 
result of the fact that the super-coset space is not a symmetric space.
The WZ term effectively makes the stability group to be super-stability
group which was discussed in \cite{PLC}.

Then we have calculated the super Yangian algebra and the Serre relation
has been also shown. It remains to clarify whether the finite size effect of the string world-sheet surely breaks the Yangian symmetry or not. 
The method of \cite{kschoutens} would be favorable in this line. 
In addition, we have constructed the transfer matrix from the Lax pair. 
From the Poisson bracket of transfer matrices, 
we can read off the classical $r$-matrix satisfying the classical
Yang-Baxter equation. The classical integrability of the
AdS-superstring may survive and imply the {\it quantum} integrability.
It is also expected that the lattice quantization of our super Yangian
would be connected to the super Yangian symmetry in ${\cal N}=4$
SYM$_4$\,, through the procedure of \cite{MacKay2}. We hope to report in
detail on the classical and quantum integrabilities in our formulation
in the future.

It is an interesting problem to consider the semi-classical quantization
of AdS-superstring \cite{GKP2,FT} in the Roiban-Siegel formulation, by
considering several classical solutions. In fact, we can embed several
bosonic classical solution rewritten in the matrix variables \cite{ART}
into this formulation. When we construct the classical solutions in the
full analysis, it is expected to obtain the super classical spin chain
corresponding to the super spin chain in the SYM side \cite{BS}.  It is
also interesting to study the Penrose limit \cite{P} of the super
Yangian algebra. Penrose limit of the Yangian algebra is discussed in
\cite{Alday}. The resulting algebra after taking the Penrose limit is
expected to be obtained via the In$\ddot{\rm o}$n$\ddot{\rm u}$-Wigner
contraction of our super Yangian algebra, as in the case of superalgebra
between AdS and pp-wave \cite{HKS}.  The pp-wave limit of the charges in
the Roiban-Siegel formulation was demonstrated in the AdS$_2\times$ S$^2$
background in \cite{Hatsuda}.  We will report these issues in the near
future as another publication.

Obviously, it is quite important to consider the {\it quantum} super
Yangian algebra by quantizing the non-local charges by following the
original work of L$\ddot{\rm u}$scher and Pohlmeyer \cite{LP}. If the
classical Yangian symmetry survive under the quantization, we can expect
that the quantum AdS-superstring theory is drastically restricted by the
quantum Yangian and may be exactly solvable. 
The matching of the spectra and the equivalence of the integrable
structures between  the spin chain  and string theories
are confirmed up to the two-loop order \cite{ArSt} on a specific example.
In fact, the Bethe
equations for the diagonalization of the Hamiltonian of quantum strings
on AdS$_5$ $\times$ S$^5$ at large string tension are proposed in
\cite{AFS}.  
It is also interesting in the direction for studies of the
quantum AdS-superstring to investigate the beta-function and conformal
fixed point, by following the scenario recently presented by
Polyakov\cite{FP}. The beta-function of non-linear sigma model with the
super Lie group as the target space, for example in the case of
$PSL(4|4)$ is also discussed by Bershadsky, Zhukov and Vaintrob
\cite{BZV}.  In the case of $PSL(4|4)$\,, the beta-function is zero due
to the vanishing dual Coxeter number.  In preparing our work, the
relationship between the integrability of AdS-string and integrable
models were also discussed in \cite{HHWXY}.

We believe that the classical and quantum integrabilities will play an
important role in studies of the AdS/CFT correspondence at non-BPS
sectors.

\section*{Acknowledgments}

We would like to thank  H.~Fuji, K.~Kamimura, M.~Sakaguchi, Y.~Satoh,
Y.~Susaki, Y.~Takayama, D.~Tomino and A.~Yamaguchi for useful
discussion.  
We would like to thank N. Berkovits and W. Siegel for fruitful discussions
especially on the $\kappa$-symmetry and the $GL(1)$ symmetry.
We appreciate to the referee for fruitful and helpful suggestions
for improving the manuscript.  
The work of K.~Y.\ is supported in part by JSPS Research
Fellowships for Young Scientists.
\appendix

\section*{Appendix}

\section{Brief introduction to $GL(n|n)$\,, $SL(n|n)$ and $PSL(n|n)$} 

In this place we shall briefly introduce several properties of 
the super Lie group, $GL(n|n)$\,, $SL(n|n)$ and $PSL(n|n)$\,. 
The knowledge of these super groups helps us to understand 
the motivation of the Roiban-Siegel formulation and it is also
applicable to other consideration in the AdS/CFT correspondence with
the super matrix formulation (For example, see \cite{BZV, BBHZZ}.).    

The supergroups $GL(n|n)$ and $SL(n|n)$ are defined as, respectively, 
\begin{eqnarray}
&& GL(n|n) := \{M \in {\rm Mat}(n|n)~ | ~{\rm sdet}M \neq 0\}\,, \\
&& SL(n|n) := \{M \in {\rm Mat}(n|n)~ | ~{\rm sdet}M = 1\}\,.  
\end{eqnarray}
Here the  $M\in$ Mat$(n|n)$ describes a $2n\times 2n$ supermatrix:
\[
 M = \begin{pmatrix}
A & B \\
C & D 
\end{pmatrix}
\,,
\]
where $A$ and $D$ in the diagonal parts are bosonic matrices, and $B$
and $C$ in the off-diagonal parts are fermionic ones. 
The super Lie algebra of SL$(n|n)$ is  
\begin{eqnarray}
&& sl(n|n) := \{M\in {\rm Mat}(n|n)~|~{\rm STr}M = 0\}\,, 
\end{eqnarray} 
where the supertrace of $M$ is defined as 
\[
 {\rm STr}M = {\rm Tr}A - {\rm Tr}D\,. 
\]

The $SL(n|n)$ has a $U(1)$ subgroup as a non-trivial center, and so it
is not simple. A generator proportional to the identity matrix $\alpha
\cdot {\bf 1}_{2n}~(\alpha \in \mathbb{C}\,, 1_{2n} \in sl(n|n))$ is
included in the $sl(n|n)$\,. This generator gives the $U(1)$ symmetry. 
Then we can define the super Lie group $PSL(n|n) \equiv SL(n|n)/U(1)$ and it
is simple.  The $PSL(n|n)$\,, however, does not have a representation
in Mat$(n|n)$\,. Notably, the $psl(n|n)$ cannot be embedded into
$sl(n|n)$ as a subalgebra. As a comment, the above-mentioned center does
not exist in the case of $sl(m|n)~(m\neq n)$ that is simple. 

The super Lie algebra $gl(n|n)$ for the supergroup $GL(n|n)$ has a
non-degenerate metric given by $g_{ij} \equiv {\rm STr}(T_iT_j)$\,, 
where $T_i$ are the generators of $gl(n|n)$ in the fundamental
representation. There are $(2n)^2$ generators of $gl(n|n)$\,, and we can
choose as follows:
\begin{itemize}
 \item supertraceless and traceless matrices $T_a$~ 
$\bigl(a = 1,\ldots, (2n)^2 - 2\bigr)$ 
 \item the identity $I \equiv 1_{2n}$ 
and the matrix $J =$diag$(1,\ldots,1,-1,\ldots,-1)$\,. 
\end{itemize}
By combining the $T_a$ with the identity $I$\,, the $sl(n|n)$ is
generated. 
In this basis, the $gl(n|n)$ metric is given by 
\begin{eqnarray}
g_{ij} = \begin{pmatrix}
g_{ab} & 0 \\
0 & \begin{array}{cc}
0 & 2n \\
2n & 0 
\end{array}
\end{pmatrix}
\,,
\end{eqnarray}
where $g_{ab}$ is an invariant metric on $psl(n|n)$\,. 

\section{Treatment of non-ultra local terms} 
\label{app:reg}

The Poisson bracket of the AdS-string in the Roiban-Siegel formulation
contains a subtlety for the non-ultra local term (i.e., the delta
function term with a derivative) as in the principle chiral model. 

The presence of the non-ultra local term in Poisson brackets of the
currents may lead to the ambiguity of the Poisson brackets of non-local
charges. Hence, we need to determine the regularization of the charges
and the order of taking the limits, in order to obtain the well-defined
Poisson brackets \cite{LP}. The subtlety of the limits in the principle
chiral model is also discussed in \cite{LP, MacKay}. 

Here it is convenient in the parameterization of world-sheet, $-\pi \leq
\sigma \leq +\pi$\,, since the treatment of non-ultra local term can
be discussed in the same way as in principle chiral models on a flat
plane. We do not loose the generality even if we consider in this
parameterization, because this parameterization can be mapped to other one
via the translation of $\sigma$\,. For example, the shift of $\sigma$\,, 
$\sigma \rightarrow \sigma + \pi$ 
leads to the range $0 \leq \sigma \leq 2\pi$\,.
 
In general, one can regularize the integral of the charges as  
\begin{eqnarray}
&& Q_{0m}^{~~n} = \lim_{X_1,X_2\rightarrow\pi}\int^{X_1}_{-X_2}\!\!\!\! 
d\sigma\, (J^R_\tau)_{m}^{~n}(\sigma)\,, \\
&& Q_{1m}^{~~n} = \lim_{Y_1,Y_2\rightarrow\pi}\Biggl[
\int^{Y_1}_{-Y_2}\!\! d\sigma\, {J^R_\sigma}_{m}^{~n}(\sigma) 
-\int^{Y_1}_{-Y_2}\!\!\!\! d\sigma\int^{Y_1}_{-Y_2}\!\!\!\!
d\sigma'\epsilon(\sigma - \sigma')[(J^R_\tau)(\sigma),\,(J^R_\tau)(\sigma')]_{m}^{~n}\Biggr]\,, 
\end{eqnarray}
but then we have the ambiguity proportional to $\displaystyle{
\lim_{X_1,X_2\rightarrow\pi}\lim_{Y_1,Y_2\rightarrow\pi}[ \th(Y_1
- X_1) - \th(Y_2 - X_2)}]$ in the Poisson bracket of $Q_0$ and $Q_1$\,.
We can avoid this ambiguity by taking the regularization as $X_1=X_2
\equiv X$ and $X_1 = Y_2 \equiv Y$\,. In other words, we can well define
the Poisson bracket of $Q_0$ and $Q_1$ by taking this regularization. 

In addition, there is an ambiguity with respect to the order of
the limits exists in calculating the $Q_0$ term in the expression of
$Q_2$\,, (\ref{Q2})\,, and we may
take three kinds of limits: 1) $\lim_{Y\rightarrow\pi}$ before
$\lim_{Y'\rightarrow\pi}$\,, 2) $\lim_{Y'\rightarrow\pi}$ before
$\lim_{Y\rightarrow\pi}$\,, and 3) $Y=Y'$\,, where $Y$ and $Y'$ are
cutoffs for the left and right $Q_{1}$ in the Poisson bracket at the l.\
h.\ s.\ of (\ref{q1q1}).  It is not, however, troublesome because all
prescriptions give the same result $-4Q_0$\,, i.e., the factor $-4$ is
universal.

Finally, let us consider the subtlety that appears in the calculation of
the Serre relation. In that time we need to take the limit
$\displaystyle{\lim_{X\rightarrow\pi}}$ before
$\displaystyle{\lim_{Y\rightarrow\pi}}$\,.  If we take limits inversely,
then the additional terms, proportional to $Q_0$\,,
\[
 -4 \delta_l^{q}\left(
\delta_m^{k}Q_{0p}^{~~n} - \delta_p^{n}Q_{0m}^{~~k}
\right) + 4 \delta_p^{k}\left(
\delta_m^{q}Q_{0l}^{~~n} - \delta_l^{n}Q_{0m}^{~~q}
\right)\,,
\]
appear and so the Serre relation is not satisfied.  Thus, it is
necessary to determine the order of the limits as above for the
well-defined Yangian algebra. In other words, the Serre relation (namely
the closure of the Yangian algebra) leads to the definite calculus of the
Poisson bracket without imposing some artificial prescription by hand.
The relationship of the Serre relation and the definition of Poisson
bracket is discussed by MacKay \cite{MacKay} in the case of a principle
chiral model. 

As a matter of course, the regularization prescription is not modified
when we consider the super matrix case.

\end{document}